\documentstyle[12pt,aps,fleqn,preprint]{revtex}
\begin{document}
\vskip 2cm
\preprint{$\hbox to 5 truecm{\hfill Alberta-Thy-22-94}\atop
{\hbox to 5 truecm{\hfill gr-qc/9406037}\atop
\hbox to 5 truecm{\hfill June 1994}}$}
\title{Why the Entropy of a Black Hole is $A/4$?}
\author{Valeri P. Frolov\thanks{Electronic address:
frolov@phys.ualberta.ca} }
\maketitle
\noindent
\vspace{.3cm}
{\centerline {\small{\em CIAR  Cosmology Program;  Theoretical
Physics
Institute,}}}
\\
{\centerline{\small{\em University of Alberta, Edmonton, Canada T6G
2J1;}}
\\
{\centerline{\small{\em P.N.Lebedev Physics Institute,  Moscow,
Russia}}

\vspace{1cm}

\begin{abstract}
\baselineskip=1.5em
A black hole considered as a part of a thermodynamical system
possesses the
Bekenstein-Hawking entropy $S_H =A_H
/(4l_{\mbox{\scriptsize{P}}}^2)$,
where   $A_H$   is   the   area   of   a   black   hole   surface
and
$l_{\,\mbox{\scriptsize{P}}}$   is   the   Planck length. Recent
attempts to
connect this entropy with dynamical degrees of freedom of a black
hole
generically did not provide the universal mechanism which allows one
to obtain
this exact value. We discuss the relation between the 'dynamical'
contribution
to the entropy and $S_H$, and show that the universality of $S_H$ is
restored
if one takes into account that the parameters of the internal
dynamical degrees
of freedom as well as their number depends on the black hole
temperature.
\end{abstract}

\vspace{4cm}

\noindent
{\centerline{PACS numbers: 04.60.+n, 03.70.+k, 98.80.Hw}

\baselineskip6.8mm

\newpage
\hspace{\parindent}

According to the  thermodynamical analogy in  black hole physics,
the
entropy of a black hole  in the Einstein theory of gravity equals
$S_H =A_H
/(4l_{\mbox{\scriptsize{P}}}^2)$,
where   $A_H$   is   the   area   of   a   black   hole   surface
and
$l_{\,\mbox{\scriptsize{P}}}=
(\hbar G/c^3)^{1/2}$   is   the   Planck length
\cite{Beke:72,Beke:73}.
The calculations  in the framework of the Euclidean approach
initiated by
Gibbons and
Hawking \cite{GiHa:76,Hawk:79} relate this quantity with the
tree-level
contribution of the gravitational action, namely the action of the
Euclidean
black hole instanton. In this approach the entropy of a black hole
has pure
topologival origin, and it remains unclear whether there exist any
real
dynamical degrees of freedom which are responsible for it. The
problem of the
dynamical origin of the black hole entropy was intensively discussed
recently
(see
e.g.,\cite{Hoof:85,FrNo:93,CaTe:93,SuUg:94,Magg:94,GaGiSt:94,BaFrZe:94
}}).
 The basic idea which was proposed is to relate the dynamical degrees
of
freedom of a black hole with its quantum excitations. This idea has
different
realizations.
In particular, it was proposed to identify the dynamical degrees of
freedom of
a black hole with the states of all fields (including the
gravitational one)
which are located inside the black hole \cite{FrNo:93,BaFrZe:94}. By
averaging
over states located outside the black hole one generates the density
matrix of
a black hole and can calculate the corresponding entropy $S_1$. The
main
contribution to the entropy is given by inside modes of fields
located  in the
very close vicinity of the horizon. It appears that the so defined
$S_1$ is
divergent. It was argued that quantum fluctuations of the horizon may
provide
natural cut-off and make $S_1$ finite. Simple estimations
\cite{FrNo:93} of the
cut-off parameter show that $S_1 \approx S_H$ .  The generic feature
of this as
well as other 'dynamic' approaches is that  the entropy of a black
hole arises
at the one-loop level. The relation between the 'topological'
(tree-level)
calculations and one-loop calculations based on the counting
dynamical degrees
of freedom of a black hole remains unclear. In particular, $S_1$
depends on the
number and characteristics of the fields, while $S_H$ does not. What
is the
general mechanism which provides the universal relation of the
dynamically
defined entropy with the universal thermodynamical value $S_H$?
The aim of this paper is to give a simple explanation of this puzzle.

A black hole considered as a part of a thermodynamical system
possesses a
remarkable property: its properties (size, gravitational field and so
on) are
determined only by one parameter (mass $M$), which in its turn in a
state of a
thermal equilibrium is directly connected with the temperature of the
system.
By varying the temperature one at the time inevitably changes all the
internal
parameters of the system. We show  that namely this property together
with
scaling properties of the free energy results in the iniversality of
the
expression for $S_H$.

We illustrate the basic idea by considering  a simple thermodynamical
system
described by the Hamiltonian
\begin{equation}\label{2}
\hat{H}=E_0
+\hat{H}^{(N)},\hspace{1cm}\hat{H}^{(N)}=\sum_{i=1}^{N}{\hat{H}_i}
{}.
\end{equation}
We suppose that $\hat{H}_i$ is a Hamiltonian of an oscillator of
frequency
$\omega_i$.  The free energy of such a system is
\begin{equation}\label{3}
F=E_0 +T \sum_{i=1}^{N}\ln [1-\exp (-\varepsilon_i /T )],
\end{equation}
where $\varepsilon_i =\hbar\omega_i$, while the entropy reads
\begin{eqnarray}\label{4}
S=-\partial_T F =\sum_{i=1}^{N}s( \varepsilon_i /T ), \hspace{.4cm}
s(x)={x \over {\exp x -1}}-\ln [1-\exp (-x )] .
\end{eqnarray}
For a system of $N$ identical oscillators $S=Ns(\varepsilon /T )$.
The
variation of the free energy reads
\begin{equation}\label{5}
dF =-SdT +\overline{\left( {\partial
{\hat{H}}
\over {\partial \lambda}}\right) }d\lambda.
\end{equation}
where $\lambda$ denotes  additional parameters on which the
Hamiltonian
depends. In the case when the internal parameters $\lambda$ are
uniquely
defined by the temperature $T$  (as it happens for black holes) one
has
\begin{eqnarray}\label{6}
dF&=&-\tilde{S} dT \\
\tilde{S} &=& S-{d{E_0} \over dT}-{d{N} \over dT} \bar{\varepsilon}
-N{d{\bar{\varepsilon}} \over dT} , \label{7}
\end{eqnarray}
where $\bar{\varepsilon}$ is the average energy of the oscillator of
frequency
$\omega(T)$ at the temperature $T$.  It is evident that for such a
system the
entropy $S$ determined by its dynamical degrees of freedom differs
from
$\tilde{S}$. For convenience in order to distinguish between $S$ and
$\tilde{S}$ we call the former 'dynamical' and the latter
'thermodynamical'
entropy.

The above consideration being applied to the case of a black hole
indicates
that generally one cannot expect  that the black-hole's
'thermodynamical'
entropy $S_H$ coincides with the 'dynamical' entropy  $S$, obtained
by summing
the contributions to the entropy of all internal degrees of freedom
of the
black hole. In order to obtain $S_H$ one must add to $S$ the term
defined by
the tree-level free energy $F_0$ as well as the terms connected with
the
dependence of the energy and number of states on the temperature.
The
remarkable fact is that the additional terms which arise due to the
dependence
of the geometry of a black hole and the number of states of quantum
fields
inside it on the temperature, exactly compensate the main "dynamical'
contribution $S$ of the quantum field to the entropy. As the result
of this
compensation the leading term $S$ simply does not contribute to
$S_H$, and
$S_H$  is  determined by the dependence of the mass (and hence the
energy) of a
black hole on the temperature (i.e., by the tree-level free energy
$F_0$).
Because different fields contribute to $S$ additively it is
sufficient to prove
this for one of the fields.

Consider a static black hole located inside a spherical cavity of
radius $r_B$.
We suppose that the boundary surface has temperature $T_B\equiv
\hbar\beta_B^{-1}$. The black hole will be in equilibrium with the
thermal
radiation inside the cavity if it has mass $M$ such that
$\beta_B=4\pi r_+
(1-r_+ /r_B )^{1/2}$, where $r_+ =2M$ is the gravitational
radius\cite{ftn0}.
This relation can be used to express $r_+$ as the function of
$\beta_B$ and
$r_B$ \ : $r_+ =r_+ (\beta_B ,r_B)$. The equilibrium is stable if
$r_B >3r_+
/2$. The tree-level contribution of the black hole to the free energy
of the
system  is \cite{York:86,BrBrWhYo:90}
\begin{equation}\label{1}
F_0 =\hbar {\cal F}_0 =\hbar r_B \left( 1-\sqrt{1-r_+ /r_B}\right)
-\pi
r_+^2\beta_B^{-1} .
\end{equation}
The tree-level contribution $\tilde{S}_0$ to the entropy $S_H$ of a
black hole
defined as $\tilde{S}_0 =-\partial_{T_B}F_0\equiv
\hbar^{-1}\beta_B^2\partial_{\beta_B}F_0$ is $S_0 =\pi r_+^2$($\equiv
A/4l_P^2\equiv S_H$). Besides this tree-level contribution there are
also
one-loop contributions directly connected with dynamical degrees of
freedom of
the black hole, describing its quantum excitations. We consider them
now in
more details.

For simplicity we consider a contribution  ${\mbox{\boldmath $F_1$}}$
of
massless conformal invariant fields to the free energy of a black
hole. It is
possible to show \cite{BaFrZe:94} that this contribution is directly
connected
with the renormalized value ${\mbox{\boldmath
$\Gamma$}}^{\mbox{\tiny{ren}}}$
of the Euclidean effective action ${\mbox{\boldmath $\Gamma$}} ={1
\over 2}
\mbox{Tr} \ln D$ for the corresponding field $\varphi$ obeying the
field
equation $D\varphi =0$ on a manifold with the metric
\begin{equation}\label{8}
ds^2=Bd\tau^2 +B^{-1}dr^2 +r^2 d\Omega^2 .
\end{equation}
Here $B=1-r_+/r$ and $d\Omega^2$ is a line element on the unit
sphere. Namely
one has ${\mbox{\boldmath $F_1$}} =\beta^{-1}{\mbox{\boldmath
$\Gamma$}}^{\mbox{\tiny{ren}}}$. The effective action is to be
calculated on
the manifold periodic in Euclidean time $\tau$ with the period
$\beta$.

The renormalized free energies ${\mbox{\boldmath $F_1$}}$ for two
conformally
related static spaces $\bar{g}_{\mu\nu}=e^{-2\omega} g_{\mu\nu}$  are
connected
as follows \cite{DoKe:78}
\begin{eqnarray}\label{9}
{\mbox{\boldmath $F_1$}} [g]&=&{\mbox{\boldmath $F_1$}}[\bar{g}]+
\Delta
{\mbox{\boldmath $F_1$}} [\omega ,g] .
\end{eqnarray}
Here  $\Delta {\mbox{\boldmath $F_1$}} [\omega ,g] =a{\cal
A}[\omega,g]+b{\cal
B}[\omega,g]+c{\cal C}[\omega,g] $,
\begin{eqnarray}\label{9b}
\hspace{-1cm}{\cal A}[\omega ,g]&=&\int{ d^{3}x g^{1 \over 2}}\left\{
\omega C_{\alpha\beta\gamma\delta} C^{\alpha\beta\gamma\delta}
+\frac{2}{3}
\left[
R+3(\Box \omega -\omega ^\sigma \omega _\sigma)
\right]
(\Box \omega -\omega ^\sigma \omega _\sigma)
\right\} ,\\
\hspace{-1cm}{\cal B}[\omega,g]&=&\int{ d^{3}x g^{1 \over 2}}
\left\{
\omega {}^{\ *}\! R_{\alpha\beta\gamma\delta} {}^{\ *}\!
R^{\alpha\beta\gamma\delta} +4R_{\mu\nu} \omega ^\mu \omega
^\nu-2R\omega
^\sigma \omega _\sigma +2 (\omega ^\sigma \omega _\sigma)^2-4\omega
^\sigma
\omega _\sigma \Box \omega
\right\}  ,\\
\hspace{-1cm}{\cal C}[\omega,g]&=&\int{ d^{3}x g^{1 \over 2}}
\left\{
\left[
R+3(\Box \omega -\omega ^\sigma \omega _\sigma)
\right]
(\Box \omega -\omega ^\sigma \omega _\sigma)
\right\} ,
\end{eqnarray}
and the coefficients $a$, $b$, and $c$ are the coefficients of
conformal
anomalies
\begin{eqnarray}\label{9c}
a&=&\frac{\hbar}{23040 \pi^2}\left[ 12h(o)+18h(1/2)+72h(1)\right] ,\\
b&=&\frac{\hbar}{23040 \pi^2}\left[ -4h(0)-11h(1/2)-124h(1)\right]
,\\
c&=&\frac{\hbar}{23040 \pi^2}\left[ -120h(1)\right] .
\end{eqnarray}
Here $h(s)$ is the number of polarizations of spin $s$. The term
$\Delta
{\mbox{\boldmath $F_1$}} [\omega ,g] $ evidently does not depend on
the
temperature $\beta$.

The useful expression can be obtained for ${\mbox{\boldmath $F_1$}}
[g]$ by
applying the formula (\ref{9}) to the particular case when
$\omega={1 \over
2}\ln B$ and  $\bar{g}_{\mu\nu}=e^{-2\omega} g_{\mu\nu}$ is the
ultrastatic
metric. In the high temperature limit the leading terms of the
renormalized
free energy ${\mbox{\boldmath $F_1$}}[\bar{g}]$ in the ultrastatic
space
$\bar{g}$ reads
\begin{equation}\label{9d}
{\mbox{\boldmath $F_1$}}[\bar{g}]  =  -\hbar
\frac{\alpha}{\beta^4}\int{d^3 x
\bar{g}^{1/2} }+\ldots,
\end{equation}
where $\alpha\equiv \frac{\pi^2}{90}\left[
h(0)+\frac{7}{8}h(1/2)+h(1)\right]
(=32\pi^4 (a+b-c/2))$.
 For the conformal massless scalar field  ($D=-\Box +\frac{1}{6}R$)
the
renormalized free energy ${\mbox{\boldmath $F_1$}}[\bar{g}]$ in the
ultrastatic
space $\bar{g}$ allows the following representation \cite{DoKe:78}
\begin{eqnarray}\label{11}
\hbar^{-1}{\mbox{\boldmath $F_1$}}[\bar{g}] & = & -\frac{\pi^2}{90}
\frac{\bar{c}_0}{\beta^4}
-\frac{1}{24}\frac{\bar{c}_1}{\beta^2}
-\overline{tr}_3 \overline{\zeta}'_3 (0,\infty )\frac{1}{2\beta}
-\frac{\bar{c}_2 }{16\pi^2} \left[ \gamma+\ln \left(
\frac{\beta}{4\pi} \right)
\right]
\nonumber \\
	& -& \frac{\pi^{3/2}}{16}\sum_{l=3}^{\infty}
\bar{c}_l \Gamma (l-3/2) \zeta_R (2l-3) \pi^{-2l}\left(
\frac{\beta^2}{4}\right) ^{l-2}  .
\end{eqnarray}
Here $\bar{c}_l=\int{d^3x \bar{g}^{1/2}\bar{a}_l (x,x)} $, and the
functions
$\bar{a}_l$ are the Minakshisundaram-DeWitt coefficients for the the
ultrastatic metric $\bar{g}$. The analogous representation is also
valid for
non-zero spins. For the conformal massless field   $\bar{a}_0 =1$ and
$\bar{a}_1 =0$.

The variation of ${\mbox{\boldmath $F_1$}}[{g}]$ with respect to the
mertic $g$
gives the renormalized stress-energy tensor $T_{\mu}^{\nu}$. For
$\beta=\beta_H$ (i.e., for a quantum field is in the Hartle-Hawking
state) this
tensor remains finite at the horizon (at $r=r_+$). For such a state
the
divergency of thermal contribution to  $T_{\mu}^{\nu}$ (obtained by
variation
of ${\mbox{\boldmath $F_1$}}[\bar{g}]$) is exactly compensated by
vacuum
polarization ('zero-point') contribution  (obtained by variation of
$\Delta
{\mbox{\boldmath $F_1$}} [\omega ,g]$). For thermal equilibrium the
change of
the temperature $\beta$ implies the change of the black hole size
$r_+$. In
such a process the 'zero-point' contribution (depending on $r_+$) is
changed in
such a way that it all the time exactly compensates the divergencies
at the
horizon of the termal contribution (depending on temperature).

Denote by $F_1$ the on-shell value of  ${\mbox{\boldmath
$F_1$}}[{g}]$, i.e.
the value of this functional calculated on the manifold with the
Euclidean
metric (\ref{8}) with the period $\beta$ in the Euclidean time.
($\beta=\hbar/T$,  where $T$ is the temperature of the system
measured at
infinity).
In general case the free energy $F_1$ contains volume divergences.
The leading
divergent near the horizon $r=r_+$ term is of the form
\begin{eqnarray}\label{13}
F_1\approx  -\hbar
\frac{\alpha}{4\pi^3}\frac{1}{r_+\varepsilon}
f\left(\frac{\beta_H}{\beta}\right)  ,
\end{eqnarray}
where $\varepsilon =(l/r_+)^2$ and $l$ is the proper distance from
the horizon,
and $\beta_H=2\pi r_+=\hbar/T_H$,  ($T_H$ is the black hole
temperature). The
function $f(x)$ in Eq.(\ref{13})  for large $x$ is $\sim x^{4}$ and
for $x=1$
it vanishes  $f(1)=0$.
This divergence (\ref{13}) for $\beta\ne\beta_H$ reflects the fact
that the
number of modes that contribute to the free energy and entropy is
infinitely
growing as one consider regions closer and closed to the horizon
\cite{FrNo:93}. In order to emphasize the dependence of $F_1$ on the
dimensionless cut-off parameter $\varepsilon$ we shall write
$F_1=F_1(\beta
,r_+ ,\varepsilon )$. For a black hole inside the cavity  the spatial
integration must be restricted by the cavity volume \cite{ftn1}. It
means that
$F_1$  depends also on $r_B$. We do not indicate this dependence
explicitly
because it is not important for our purpose.
For  $\beta=\beta_H$ the not only the leading but all possible volume
divergencies  disappear.  The value   $F_1 (\beta_H ,r_+ ,0)$ in this
limit is
finite because the point $r=r_+$ is regular pint of the regular
Euclidean
manifold and the renormalized effective action calculated for a part
$r\le r_B$
 of this manifold is finite.

The free energy has the same dimension as $\hbar r_+^{-1}$ and hence
it can be
presented in the form
\begin{eqnarray}\label{15}
F_1(\beta ,r_+ ,\varepsilon )\equiv r_+^{-1}\hbar {\cal F}
(r_+^{-1}\beta
,\varepsilon ) ,
\end{eqnarray}
where ${\cal F}$ is dimensionless function of two dimensionless
variables. This
property also directly follows from  the following scaling property
of the
renormalized free energy \cite{DoKe:78}
\begin{equation}\label{14}
F_1(\beta ,r_+ ,\varepsilon )=\alpha F_1(\alpha\beta ,\alpha r_+
,\varepsilon )
{}.
\end{equation}
The finiteness of  $F_1 (\beta_H ,r_+ ,0)$ and the scaling (\ref{14})
are the
properties which are not connected   with the particular choice of
the field
and remain valid for an arbitrary quantum field.

The one-loop contribution of a quantum field $\varphi$ to the entropy
is
\begin{equation}\label{16}
S_1 =\frac{1}{\hbar}\left[ \beta^2 {{\partial {F_1}}\over {\partial
\beta
}}\right] _{\beta =\beta_H} .
\end{equation}
It should be stressed that one must put $\beta=\beta_H$ only after
the
differentiation. The leading near the horizon term of $S_1$ is
\begin{eqnarray}\label{17}
S_1\approx \frac{2\alpha}{\pi^2\varepsilon} .
\end{eqnarray}
If the proper-distance cut-off parameter $l$ is of the order of
Planck length
$l_P$ then the contribution of the field to the entropy $S_1$ is of
order $S_1
\sim A/l_P^2$, where $A$ is the surface area of the black hole. In
other words
for the 'natural' choice of the cut-off parameter $l\sim l_P$ the
'dynamical'
entropy $S_1$ of a black hole is of the same order of magnitude as
the
'thermodynamical' entropy $S_H$.

We return now to our main problem: the relation between $S_1$ and
$S_H$. One
can write
\begin{eqnarray}\label{18}
dF_1\equiv -\tilde{S}_1 dT_H=-(S_1 +\Delta S )dT_H .
\end{eqnarray}
The first relation defines the one-loop contribution  $\tilde{S}_1$
to the
'termodynamical' entropy of the black hole. The additional term
\begin{eqnarray}\label{19}
\Delta S =\frac{\beta^2_H}{4\hbar\pi}\left[ {\partial {F_1}\over
\partial
{r_+}}+{\partial {F_1}\over \partial {\varepsilon}}{\partial
{\varepsilon}\over
\partial {r_+}} \right] _{\beta=\beta_H} .
\end{eqnarray}
is connected with the change of energy of  states of the field when
the
parameters of the system ($r_+$ in our case) are changed (the first
term in the
square brackets of Eq.(\ref{19})),  and with the change of the number
of states
in these process  (the second term in Eq.(\ref{19}) ). This terms are
similar
to the anologous terms arizing in Eq.(\ref{7}). It is evident that
$\tilde{S}_1$ can also be obtained by direct substitution of
$\beta=\beta_H\equiv 4\pi r_+$  into $F_1$ before its differentiation
with
respect to $\beta_H$. By using Eq.(\ref{15}) one gets
\begin{equation}\label{20}
\tilde{S}_1 =4\pi\beta^2_H{\partial \over \partial \beta_H}\left[
{\frac{\cal
F}{\beta_H}}\right] =
4\pi\left[ -{\cal F}(4\pi,\varepsilon)+{\partial {{\cal
F}(4\pi,\varepsilon
)}\over \partial \varepsilon}{\partial {\varepsilon}\over \partial
{\ln
\beta_H}}\right] .
\end{equation}
Because the  free energy $F_1$ does not contain volume divergencies
for
$\beta=\beta_H$, the quantity in the square brackets of Eq.(\ref{20})
is a
finite number. It means that  the additional contribution $\Delta S$
which
arises due to the change the parameters of the system (of the black
hole)
exactly compensates the divergent terms of $S_1$ ('dynamical'
entropy). That is
why the contribution $\tilde{S}_1$ of the quantum field $\varphi$ to
the
'thermodynamical' entropy of a black hole $S_H$ is of order of
$O(\varepsilon^0)$.  It is much smaller than $A/l_P^2$ and can be
neglected.
As the result of this compensation mechanist the dynamical degees of
freedom of
the black hole practically do not contribute to its 'thermodynamical'
entropy
$S_H$, and the letter is defined by the tree-level quantity
$\tilde{S}_0$.
It is evident that this conclusion remains valid for contribution of
any other
field.

To make the basic idea more clear  we restricted ourselves in the
above
discussion by considering a non-rotating black hole. The analysis is
easily
applied to the case of a charged rotating black hole as well as to
their
n-dimensional generalization. It is interesting that for black holes
in the
generalized gravitational theories the 'thermodynamical' $S_H$  and
'dynamical'
$S_1$ entropy may have different dependence on the mass $M$ of a
black hole. In
particular for two dimensional dilaton black hole $S_H=4\pi
M/\sqrt{\lambda}$ \
\cite{Frol:92},  while $S_1 \sim \ln \varepsilon$.

To summarize we show that there is no contradiction between
thermodynamical
definition of the entropy and its statistical-mechanical calculation
based on
the existence of the internal degrees of freedom of the black hole.
The
dynamical  degrees of freedom of the black hole are related with
possibility
for a black hole to have different internal structure for the same
external
parameters.  A black hole as a thermodynamical system is singled out
by the
property that the parameters of its internal degrees of freedom
depend on the
temperature of the system in the universal way. This results in the
universal
cancellation of all those contributions to the thermodynamical
entropy which
depend on the particular properties and number of  fields. That is
why the
'thermodynamical' entropy of  black holes in the Einstein theory is
always
$S_H$.

\vspace{.5cm}
{\bf Acknowledgements}:\ \ The author  thanks Andrei Barvinsky and
Andrei
Zelnikov  for helpful discussions.
This work was supported  by the Natural Sciences and Engineering
Research Council of Canada under the Research Grant OGP0138712.


\begin{thebibliography}{99}
\bibitem{ftn1} For a black hole  located inside a cavity  the field
$\varphi$
must obey some boundary conditions at the surface of the cavity.  For
this case
boundary terms must be also added to the expression (\ref{11}). This
modification is not important for the entropy problem. That is why we
do not
discuss these boundary terms.
\end{thebibliography}
\end{document}